\documentclass[sigconf,nonacm,10pt]{acmart}

\AtBeginDocument{%
  }

\usepackage{amsmath}

\usepackage{amssymb}
\usepackage{algorithm}
\usepackage[noend]{algpseudocode}
\usepackage{subcaption}
\usepackage{enumitem}

\usepackage{multirow}
\usepackage{booktabs}
\usepackage{array}
\usepackage{graphicx}
\usepackage{capt-of}
\usepackage{hhline}

\usepackage{xcolor}

\usepackage{hyperref}
\usepackage{cleveref}
\crefname{section}{§}{§§}

\newcommand{\sysname}{\textsf{FugaStream}}

\setlist[itemize]{topsep=0pt, leftmargin=*}

\begin{document}

\title{FugaStream: Pipelined Live Volumetric Streaming towards Full Exploitation of Multi-views}

\author{Edward Jusuk Kim}
\email{jusukkim@snu.ac.kr}
\affiliation{%
  \institution{Seoul National University}
  \city{Seoul}
  \country{Republic of Korea}
}

\author{Huijeong Choe}
\email{huijeongchoe@snu.ac.kr}
\affiliation{%
  \institution{Seoul National University}
  \city{Seoul}
  \country{Republic of Korea}
}

\author{Minjae Lee}
\email{minjaelee@snu.ac.kr}
\affiliation{%
  \institution{Seoul National University}
  \city{Seoul}
  \country{Republic of Korea}
}

\author{Chanjeong Park}
\email{cjpark99@snu.ac.kr}
\affiliation{%
  \institution{Seoul National University}
  \city{Seoul}
  \country{Republic of Korea}
}

\author{Kyunghan Lee}
\email{kyunghanlee@snu.ac.kr}
\affiliation{%
  \institution{Seoul National University}
  \city{Seoul}
  \country{Republic of Korea}
}

\renewcommand{\shortauthors}{Kim et al.}

\renewcommand{\shortauthors}{Kim et al.}

\begin{abstract} %
Live volumetric streaming has become the cornerstone of next-generation immersive applications, enabling users to experience six degrees of freedom (6-DoF). To support interactivity, live volumetric streaming must reconstruct multi-view RGB-D frames into a unified 3D representation while meeting a stringent latency budget, where existing systems are forced to trade reconstruction quality for latency. In this paper, we present~\sysname, the first pipelined live volumetric streaming system that mitigates this trade-off by committing a base representation early and then incrementally merging novel late views. Enabling such functionalities introduces two fundamental challenges: (i) the uncertainty in frame availability, and (ii) the late view updates in a lightweight manner. We address these challenges through two key algorithms: (i) \textit{adaptive base commitment control}, which determines the base commitment time with the expected reconstruction gain of each missing view against the growing cost of latency, and (ii) \textit{base-referenced late view completion}, which transmits only the novel regions of late views compared to early committed base, enabling lightweight fusion at the client. We validate the effectiveness of~\sysname~on a widely used volumetric dataset and a real-world testbed, demonstrating that ~\sysname~improves SSIM by 10.9\% and IoU by 13.7\% over baselines under the same 150 ms budget.
\end{abstract}


\maketitle

\section{Introduction}
\label{sec:introduction}

\begin{figure}[t]
    \centering
    \includegraphics[width=\linewidth]{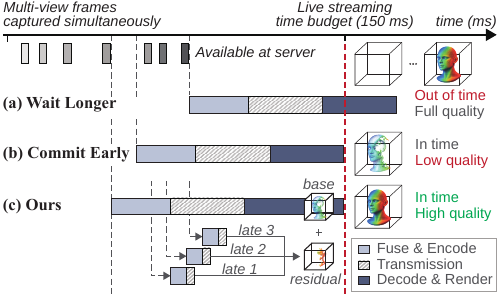 }
    \caption{Latency--quality trade-off in live volumetric streaming: (a) waiting for all views preserves quality but exceeds the latency budget, (b) committing fusion early meets the latency budget at the cost of quality, and (c) pipelined streaming exploits as many views as possible, meeting the latency budget without sacrificing quality.}
    \label{fig:fig1}
\end{figure}

With the commercialization of volumetric computing platforms ~\cite{apple_visionpro,meta_quest3,starline,codec_avatar}, live volumetric streaming has emerged as a key enabler of next-generation interactive applications, including multi-party telepresence~\cite{starline,holoportation}, immersive classroom lectures~\cite{holoclass}, and collaborative mixed-reality gaming~\cite{blocks}. In such services, users can freely move and change their viewpoint with six degrees of freedom (6-DoF), feeling as if they are physically co-located with remote participants.

To enable such immersive experiences, the core of the live volumetric streaming pipeline lies in aggregating multi-view frames and reconstructing them into a unified 3D representation (\textit{e.g.,} point cloud, voxel grid, or mesh) at the server ~\cite{fumos,livescan3d,groot,vivo,metastream,pointpresence}. In this multi-view setting, multiple RGB-D cameras observe the same object from different angles, with their fields of view partially overlapping over the object's surface. These overlapping observations allow views to compensate for one another by suppressing noise, while each distinct viewpoint captures regions occluded from the others~\cite{vivo,immerscope}. Therefore, the server must incorporate all views into 3D reconstruction for high-fidelity live volumetric streaming.

However, the requirement to aggregate all views before reconstruction conflicts with the strict latency budget of immersive applications. To preserve interactivity, the entire capture-to-display pipeline is required to meet the latency constraint of approximately 150 ms~\cite{ITU-G1010}. Within this budget, each frame must traverse multiple processing and communication stages along the end-to-end pipeline, leaving only a limited amount of time for server-side view aggregation and 3D reconstruction. Compounding this constraint, frames captured synchronously by different cameras become ready asynchronously for reconstruction, with their readiness often spread over tens of milliseconds~\cite{livo,livevv}. Such variations arise from RGB-D encoding~\cite{temporalRVL,lossless_depth}, uplink contention~\cite{metastream,livevv}, server-side pre-processing such as background and noise filtering~\cite{characterizing}, and other overheads. This readiness spread pulls the server in two opposite directions---waiting for all views can exceed the latency budget (Figure~\ref{fig:fig1}(a)), while starting reconstruction early can leave the reconstructed 3D model significantly incomplete (Figure~\ref{fig:fig1}(b)).

In this paper, our goal is to achieve a better operating point between these two extremes by starting reconstruction before every view is ready, while still exploiting late views. As illustrated in Figure~\ref{fig:fig1}(c), the server commits to reconstruction once the available views are sufficient. The resulting reconstruction forms a base representation, which is transmitted to the client without waiting for the late views. Views that become ready after the base commitment are incorporated as lightweight updates to refine the base. 

Realizing this system poses two key challenges. First, determining when the available views are sufficient for the base commitment is fundamentally challenging. Neither the reconstruction-ready time of each view nor the identity is known in advance. Waiting longer may yield only marginal quality gains, while committing early may discard views that would substantially improve the reconstruction. Second, identifying the regions of each late view that provide novel information beyond the base is nontrivial. A late view may observe many surface regions that are already represented in the base, while also revealing previously occluded regions that can improve the reconstruction. Treating the entire late view as new information therefore introduces redundant transmission and reconstruction overhead for regions already represented in the base.

To this end, we propose~\sysname, the first pipelined live volumetric streaming system, to the best of our knowledge, that mitigates the rigid dependency between reconstruction timing and view utilization.
~\sysname~consists of the following two key components:

\noindent \textbf{Adaptive base commitment control.} 
Whenever a view becomes available,~\sysname~evaluates whether to wait for additional views or commit the base reconstruction with the current set. It estimates the probability of each missing view becoming available in the near future based on recent readiness history. By weighing the expected reconstruction gain against the rapidly increasing cost of waiting,~\sysname~continues waiting only when the expected gain justifies the delay, and otherwise commits the currently available views as the base reconstruction.

\noindent \textbf{Base-referenced late view completion.} 
To address the significant cost of client-side updates, this module enables lightweight incremental refinement using intermediate fusion states. Along with the base representation,~\sysname~ transmits its underlying states, including surface distance and observation confidence. When late views become ready after base commitment, these states are used to filter out regions already represented in the base and identify only novel geometry. The extracted novel regions are then seamlessly integrated into the base at the client in a pipelined manner. By carrying the reconstruction states with the base and sending only the novel portions of late views,~\sysname~reduces update overhead, while preserving the reconstruction quality gained from late views.

We summarize our contributions as follows.
\begin{itemize}[topsep=0pt]
    \item We propose~\sysname, the first pipelined live volumetric streaming system that eliminates the need to wait for all views before reconstruction by incrementally refining the result as additional views become ready rather than blocking until all views are received.

    \item We design adaptive base commitment control and base-referenced late view completion that adaptively decide when to commit a base reconstruction and how to refine it with late-arriving views, allowing~\sysname~to exploit as many views as the latency budget permits.

    \item We evaluate~\sysname~on a widely used volumetric dataset as well as on a real-world testbed. Compared with existing approaches,~\sysname~improves SSIM and IoU by 10.9\% and 13.7\%, respectively, at the latency budget of 150 ms.
\end{itemize}

\section{Background}
\label{sec:background}

\begin{figure}[t]
    \centering
    \includegraphics[width=\linewidth]{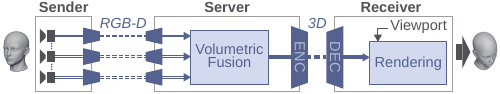}
    \caption{Conventional live volumetric streaming pipeline. Multiple RGB-D views are transmitted to the server, fused into a 3D representation, encoded, and delivered to the receiver for viewport rendering.}
    \label{fig:volumetric_streaming}
\end{figure}

A volumetric streaming pipeline generally consists of three stages: (i) multiple RGB-D camera capture, (ii) server-side 3D reconstruction, and (iii) client-side rendering, as shown in Figure~\ref{fig:volumetric_streaming}. To ensure consistency across views, RGB-D cameras at different viewpoints capture the scene simultaneously. We refer to the synchronized multi-view frames associated with the same capture timestamp as a \textit{frame set}. Upon receiving the frame set, the server removes background and noise, and merges the processed views into a unified 3D representation. The reconstructed 3D data is then transmitted to the client to render the object from the user's current viewpoint.

\begin{figure*}[t]
    \centering

    \begin{minipage}[t]{0.24\textwidth}
        \vspace{0pt}
        \centering

        \includegraphics[width=\linewidth]{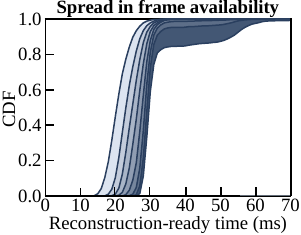}
        \vspace{-0.18in}
        \captionof{figure}{Spread in frame availability across views.}
        \label{fig:frame_spread}
    \end{minipage}%
    \hfill%
    \begin{minipage}[t]{0.49\textwidth}
        \vspace{0pt}
        \centering

        \vspace{0.1in}
        \includegraphics[width=\linewidth]{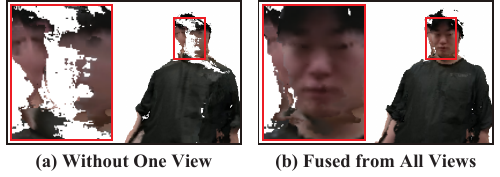}
        \vspace{-0.18in}
        \captionof{figure}{Qualitative example of impact of a missing view.}
        \label{fig:quality_drop}
    \end{minipage}%
    \hfill%
    \begin{minipage}[t]{0.24\textwidth}
        \vspace{0pt}
        \centering
        
        \includegraphics[width=\linewidth]{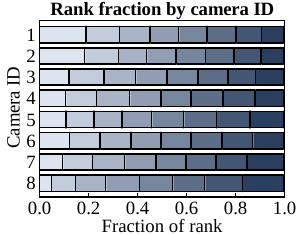}

        \captionof{figure}{Distribution of fusion-ready rank across cameras.}
        \label{fig:cam_id}
    \end{minipage}
    \vspace{-0.1in}
\end{figure*}

To support rendering from arbitrary viewpoints, the 3D reconstruction must combine measurements from all views into a coherent surface representation without holes or discontinuities between views. For this purpose, truncated signed distance function (TSDF) fusion is widely used in volumetric reconstruction~\cite{sdf,kinectfusion}. A TSDF represents 3D space as a voxel grid, where each voxel stores a TSDF value and an associated confidence weight. The TSDF value encodes the signed distance from the voxel to the observed surface along the camera ray, while the confidence weight determines its contribution during fusion. For each camera view, a voxel is projected onto the corresponding depth map, and its depth is compared with the observed surface depth to obtain the signed distance. Voxels in front of the surface take positive values and those behind take negative values, with the distance truncated to a bounded range around the surface. When multiple views observe the same region, their TSDF values are combined voxel by voxel using a weighted average. The reconstructed surface is then obtained from the zero-crossing of the fused TSDF field.

An important property of TSDF fusion is that each view is not represented as separate 3D data, but is accumulated into a shared voxel grid. We exploit this shared representation to support base-referenced late view completion---a base model can first be constructed from the views available at commitment time, and late views can subsequently update the same voxel grid as they arrive, rather than requiring reconstruction from scratch.

\section{Motivation}
\label{sec:motivation}

In this section, to rigorously investigate the characteristics of live volumetric streaming, we conduct an empirical analysis using a real-world testbed. Our testbed configuration is summarized in Table~\ref{tab:real-world_testbed}. Each camera captures RGB-D frames at 30 fps, \textit{i.e.}, with a frame period of 33 ms, while a synchronization hub aligns the capture instants across cameras within 1 ms. Each camera node is equipped with an NVIDIA Jetson Nano for local frame processing and transmission. As discussed in~\cref{sec:introduction}, when views in a frame set do not become ready together, the server faces two extremes: waiting longer or committing early. We first quantify the spread in frame availability observed in our testbed, and then evaluate the impact of each choice on latency and reconstruction quality.

\begin{table}[t]
    \centering
    \small
    {\begin{tabular}{c|l}
        \specialrule{1pt}{0pt}{0pt}
        \multirow{2}{*}{\textbf{Cameras}}
            & 8 Orbbec Femto Mega RGB-D cameras~\cite{orbbec-femto} \\
            & NVIDIA Jetson Nano \\ \hline
        \multirow{3}{*}{\textbf{Server}}
            & AMD Ryzen 9 7950X \\
            & 32 GB DDR5 \\
            & NVIDIA GeForce RTX 4090 \\ \hline
        \textbf{Client}
            & NVIDIA Jetson Orin NX \\ 
        \specialrule{1pt}{0pt}{0pt}
    \end{tabular}}
    \vspace{0.05in}
    \caption{Specifications of real-world testbed. All components are connected over 1 Gbps Ethernet.}
    \label{tab:real-world_testbed}
\end{table}

\noindent\textbf{Large spread in view availability.} Figure~\ref{fig:frame_spread} shows the CDFs of the fusion ready times of the eight views, ranked by the order in which they become ready within each frame set. Over a video sequence comprising 5,180 frame sets, we measure the elapsed time from capture to fusion readiness at the server for each view. At p99, the first-ready view becomes available in approximately 30 ms, whereas the last-ready view takes more than 65 ms. Given the 33 ms frame period, views from the next frame set may already become available while the current frame set is still incomplete, potentially causing pipeline stalls or frame drops if reconstruction waits for every view. In the worst case, the last-ready time takes up to 138 ms, consuming nearly 92\% of the 150 ms end-to-end latency budget before reconstruction even begins.

\noindent\textbf{Quality degradation from a missing view.} The defining characteristic of volumetric streaming lies in supporting 6-DoF, requiring sufficient surface coverage for rendering from arbitrary viewpoints. A missing view can therefore leave unobserved regions that become exposed as the user moves. Figure~\ref{fig:quality_drop} shows that omitting even a single view can significantly degrade reconstruction quality. In this example, the absence of only one view introduces substantial geometric distortion and leaves large portions of the surface incomplete. Such degradation is particularly problematic for volumetric streaming, where users can freely change their viewpoint and directly expose the affected regions.

These observations show that neither waiting for all views nor committing early without late views is desirable for live volumetric streaming. Motivated by these two requirements, we propose a pipelined streaming system in which the server reconstructs and delivers a base representation using the views already available, while late views are subsequently incorporated into the base at the client side as they become ready. This allows reconstruction to proceed without waiting for the complete frame set, while still leveraging late views for subsequent refinement.

\section{Challenges}
\label{sec:challenges}

We establish a pipelined volumetric streaming system that commits a base early and incorporates late views as they become ready. However, achieving this presents fundamental challenges that do not arise in conventional live volumetric streaming. In this section, we characterize these challenges.

\noindent\textbf{Uncertainty in view availability.} A static commitment decision is ineffective because view readiness is not consistent across frame sets. To examine this variability, Figure~\ref{fig:cam_id} shows, for each view, the fraction of frame sets in which its frame becomes fusion-ready at each rank from 1 to 8. As shown in the figure, no single view dominates the last-ready rank. Every view becomes the last-ready one in a non-negligible fraction of frame sets, leaving no persistent late view that can be identified through offline profiling. Consequently, a fixed commitment control discards different views from one frame set to the next. Since each view covers a different portion of the surface, the missing regions also change across frames, appearing to the user as temporally unstable holes or flicker. Simply extending the wait time does not resolve the problem: due to the varying availability spread (\cref{sec:motivation}), it possibly causes idle waiting in some frame sets while still failing to include long-tail views in others. The commitment decision must therefore adapt to this uncertainty observed for each frame set at runtime.%

\begin{figure}[t]
    \centering
    \includegraphics[width=\linewidth]{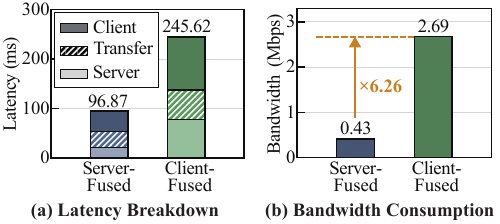}
    \caption{Cost of transmitting and fusing multi-view observations independently at the client versus fusing them at the server.}
    \label{fig:serverclient}
\end{figure}

\noindent\textbf{Cost of incorporating late views.} Incorporating late views at fine granularity can diminish the latency and bandwidth benefits gained from early commitment. A straightforward design is to encode and transmit each late view as soon as it becomes ready, allowing the client to incrementally fuse the received views. However, multi-view observations contain substantial overlap, so treating each view as an independent update can repeatedly transmit and fuse information already captured by other views. Figure~\ref{fig:serverclient} quantifies this overhead by comparing two designs. \textit{Server-Fused} fuses all eight views at the server and sends the fused representation to the client, whereas \textit{Client-Fused} transmits each view independently and performs fusion at the client. The latency measurement excludes server-side waiting and spans from the start of fusion until the frame is ready for rendering, while bandwidth measures server-to-client traffic. As shown in the figure, fine-grained client-side fusion incurs substantially higher latency and bandwidth than server-side fusion. The system must therefore incorporate the useful contribution of late views without repeatedly transmitting and fusing information already contained in the base.

Taken together, these observations show that simply decomposing volumetric streaming into multiple stages is insufficient. Because view availability varies across frame sets, the system must adapt its commitment decision to the readiness observed at runtime. At the same time, processing every late view independently incurs redundant transmission and fusion, increasing both latency and bandwidth consumption. Therefore, the system must incorporate late views while limiting this additional cost. In the next section, we present our system design for addressing these challenges.

\section{System Design}
\label{sec:system_design}

\subsection{System Overview}
\label{subsec:system_overview}
\begin{figure}[t]
    \centering
    \includegraphics[width=\columnwidth]{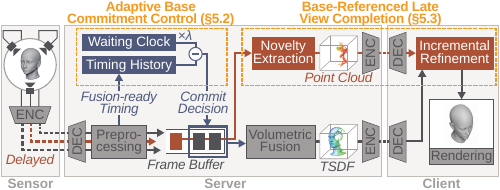 }
    \vspace{-0.2in}
    \caption{System overview of~\sysname, a pipelined volumetric streaming system that commits early and then merges late views to improve latency--quality trade-off.}
    \label{fig:system_overview}
    \vspace{-0.25in}
\end{figure}

The overall system architecture of \sysname~is illustrated in Figure~\ref{fig:system_overview}. In this system, multiple RGB-D frames are transmitted to the server with varying delays. To commit the base, the server estimates when each view will become available and adaptively adjusts its waiting time (\cref{subsec:base_commitment_control}). The server then commits the base reconstruction and transmits only its sparse surface-band voxels. This reduces both the transmission bandwidth and the client-side computation. For late views after the base transmission, the server extracts only the novel regions that the base does not cover sufficiently (\cref{subsec:late_view_completion}). The client merges these late views into the base and incrementally updates the 3D representation within the live streaming latency requirement.

To realize this, we define the following design goals:
\begin{itemize}[topsep=0pt]
    \item \textbf{End-to-end latency for interactivity.} We aim to deliver a base representation within the application’s end-to-end latency budget. To this end, the server decides how long it can wait before fusion.
    
    \item \textbf{Late view utilization.} At the same time, useful geometric contributions from late views should be preserved as they become available. Our system therefore keeps the base representation refinable and integrates only novel regions from late views with minimal overhead.
\end{itemize}

\subsection{Adaptive Base Commitment Control}
\label{subsec:base_commitment_control}
To construct a base representation, we introduce an adaptive base commitment control (ABCC) algorithm that determines when to commit the base reconstruction. Algorithm~\ref{alg:commit} summarizes the workflow of ABCC. Upon each view $i \in \{1,...,N\}$ becoming fusion-ready, ABCC evaluates if an additional wait time $\Delta^*$ for views that have not yet become ready $\mathcal{M}_g$ is acceptable. Each view that becomes available within $\Delta^*$ is added to $\mathcal{R}_g$, the set of views to be included in the base. ABCC derives $\Delta^*$ from two factors: (i) the coverage gain obtained by waiting for the views in $\mathcal{M}_g$, and (ii) the waiting cost which grows exponentially as time elapses.

\noindent\textbf{Coverage gain.} As discussed in~\cref{sec:motivation}, the timing and the ordering of each view's availability vary significantly across frame sets. To adapt to such variations, ABCC maintains a sliding window over the recent fusion-ready time histories of each camera, $\{\mathcal{H}_i\}$, covering the last 150 frame sets. The empirical CDF of fusion-ready times for each frame set, $\{F_i\}$, is then derived from this window. Views that arrive after the commitment are still recorded, and those that never arrive are recorded as $+\infty$. Given the current time $t$, the conditional probability $p_i(t, \Delta)$ that a view $i\in\mathcal{M}_g$ becomes ready after waiting for an additional $\Delta$ is calculated as follows:

\begin{algorithm}[t]
\caption{Adaptive Base Commitment Control}
\label{alg:commit}
\begin{algorithmic}[1]

\Statex \textbf{Input:}
Per-view fusion-ready histories $\{\mathcal{H}_i\}_{i=1}^{N}$

\Statex \textbf{Output:}
Committed base view set $\mathcal{R}_g \subseteq \{1,\ldots,N\}$

\For{each frame set $g$}

    \State $\{F_i\} \gets \textsc{EstimateCDF}(\{\mathcal{H}_i\})$

    \State $\mathcal{R}_g \gets \emptyset$
    \Comment{Reset candidate base set}

    \While{$g$ is uncommitted}

        \State Wait for \textsc{ViewReady}$(i,t)$ or timeout

        \If{timeout expires}
            \State \textbf{commit} $\mathcal{R}_g$; \textbf{break}
        \EndIf

        \State $\mathcal{R}_g \gets \mathcal{R}_g \cup \{i\}$

        \If{$|\mathcal{R}_g| = N$}
            \State \textbf{commit} $\mathcal{R}_g$; \textbf{break}
        \EndIf

        \State $\mathcal{M}_g
        \gets \{1,\ldots,N\}\setminus\mathcal{R}_g$
        \Comment{Uncommitted view set}

        \State $\Delta^*
        \gets \textsc{SelectWait}(t, \{F_i\}_{i\in\mathcal{M}_g})$

        \If{$\Delta^* = 0$}
            \State \textbf{commit} $\mathcal{R}_g$; \textbf{break}
        \Else
            \State Set timeout at $t+\Delta^*$
        \EndIf

    \EndWhile

    \State Update $\{\mathcal{H}_i\}$
    \Comment{Record fusion-ready times}

\EndFor

\end{algorithmic}
\end{algorithm}

\begin{equation} 
    p_i(t,\Delta) = \frac{F_i(t+\Delta)-F_i(t)} {\max(1-F_i(t),\epsilon)}, 
    \label{eq:conditional-arrival} 
\end{equation} 
\noindent where a small constant $\epsilon$ avoids numerical instability. The coverage gain $G_t(\Delta)$ obtained by waiting for $\Delta$ is then
\begin{equation}
    G_t(\Delta) = \frac{1}{N}\sum_{i\in\mathcal{M}_g}p_i(t, \Delta).
    \label{eq:coverage_gain}
\end{equation}
Since each view contributes surface observations that no other view provides, a larger $G_t$ indicates a more complete base at commitment.

\noindent\textbf{Waiting cost.} Since $G_t$ increases monotonically with $\Delta$, the coverage gain alone always favors waiting longer. However, waiting is not free: every additional wait delays base commitment and directly increases end-to-end latency. Moreover, the same waiting interval $\Delta$ incurs a larger penalty as the current time $t$ increases, making both how long and when to wait important. We therefore formulate the waiting cost from $t$ to $t + \Delta$ as
\begin{equation} 
    C_t(\Delta) = e^{(t+\Delta)/\tau} - e^{t/\tau}, 
    \label{eq:latency-cost} 
\end{equation} 
\noindent where $\tau$ controls how rapidly the cost grows with time. As base commitment is delayed, the increasing cost discourages further waiting unless the coverage gain is sufficiently large, thereby helping guarantee the latency budget.

\noindent\textbf{Commitment decision.} Combining Eq.~\ref{eq:coverage_gain} and Eq.~\ref{eq:latency-cost}, ABCC defines the objective for waiting an additional $\Delta$ as:
\begin{equation}
    J_t(\Delta) = G_t(\Delta) - \lambda C_t(\Delta),
    \label{eq:vow}
\end{equation}
\noindent where $\lambda > 0$ is a hyperparameter that trades off the coverage gain against the waiting cost. ABCC then determines the waiting time $\Delta^*$ as:
\begin{equation}
    \Delta^* = \mathop{\arg\max}_{\Delta} J_t(\Delta ).
\end{equation}
If $\Delta^* = 0$, \textit{i.e.}, the expected coverage gain no longer outweighs the waiting cost, the system stops waiting and commits base reconstruction. Otherwise, it extends the additional wait time by $\Delta^*$. We validate the effectiveness of ABCC in~\cref{sec:evaluation}.

\subsection{Base-Referenced Late View Completion}
\label{subsec:late_view_completion}
Since the base is transmitted to the client before all views are ready, we introduce a base-referenced late view completion algorithm that incorporates the late views into the base. As noted earlier, inter-view redundancy consumes bandwidth and client-side computation without contributing new surface information. We therefore describe how the server extracts only the novel regions of each late view, and how the client merges the received regions into the base.

\noindent\textbf{Server-side novelty extraction.} Once the base is reconstructed, redundancy arises in two places: between the base and each late view, and among the late views themselves. The server therefore extracts, for each late view, only the regions that are novel with respect to both.

First, the server removes the redundancy against the base. For an observation $\mathbf{x}$ of a late view mapped to voxel $v(\mathbf{x})$ of the base grid, we define its novelty score as:
\begin{equation}
    P_g(\mathbf{x}) = S(\mathbf{x}) \left(1-C_g(v(\mathbf{x}))\right),
\label{eq:novelty_score}
\end{equation}
\noindent where $S(\mathbf{x})$ denotes the local depth support of $\mathbf{x}$ and $C_g(v)$ the normalized coverage of voxel $v$ in the frozen base. $S(\mathbf{x})$ is computed from the consistency among neighboring depth samples to suppress inconsistent measurements. $C_g(v)$ is derived from the accumulated fusion weight $W_g(v)$ and the total number of committed views $|\mathcal{R}_g|$ as
\begin{equation}
    C_g(v) = \min\left(\frac{W_g(v)}{|\mathcal{R}_g|},1\right).
\end{equation}
$C_g(v)$ approaches $1$ when the voxel is already well covered by the base. An observation whose score falls below a threshold $\theta_p$ is regarded as redundant against the base and is discarded.

The server then removes the redundancy among the late views. For this, it maintains a transient overlap map per frame set. Whenever a voxel is selected by a late view, the map records it, and any observation of a subsequent late view that maps to the same voxel is discarded. Each voxel is thus carried by exactly one late view, and the map is discarded once the frame set is complete. The resulting set of sparse points reduces both downstream bandwidth and redundant client-side fusion computation.

\begin{figure}[t]
    \centering
    \includegraphics[width=\columnwidth]{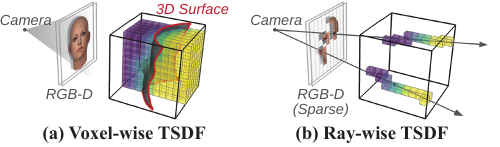 }
    \caption{Comparison of TSDF fusion paths. (a) Voxel-wise fusion projects the voxel grid onto the depth map, while (b) ray-wise fusion traverses along rays.}
    \label{fig:fusion}
\end{figure}

\noindent\textbf{Client-side incremental base refinement.} The client incorporates the late views incrementally as they arrive. As described in~\cref{sec:background}, TSDF fusion typically projects the voxel grid onto the depth map. In this voxel-wise fusion, shown in Figure~\ref{fig:fusion}(a), every voxel is updated regardless of how much the view actually observes. Repeating this for every view arrival imposes a significant computation overhead on the client.

To address this, we adopt a ray-wise fusion, presented in Figure~\ref{fig:fusion}(b). Since the server extracts only the novel regions, each late view consists of a small number of points, and updating every voxel is unnecessary. For each late view arrival, the client reconstructs the viewing ray from the camera calibration and updates only the voxels which the ray casts through. Specifically, it traverses the TSDF truncation band along the ray, updates the voxel weights, and replaces the stored color when the late view observes the surface at an angle closer to front-facing than the already stored one.

Overall, these mechanisms allow volumetric fusion to begin before all views have become available, without sacrificing the quality those views would have contributed. ABCC bounds the wait for the base view set so that fusion starts within the latency budget, while base-referenced late view completion recovers the coverage missing from the base representation. \sysname~meets the end-to-end latency budget without dropping the views that become available after commitment. We validate the effectiveness of \sysname~in~\cref{sec:evaluation}.

\section{Evaluation}
\label{sec:evaluation}

\subsection{Evaluation Setup}
\label{subsec:eval_setup}
\noindent\textbf{Network settings.} We emulate the camera-to-server network using Linux traffic control (\texttt{tc}~\cite{linux_tc}) with \texttt{netem}~\cite{netem}, configuring each link with a bandwidth of 600 Mbps and an RTT of 40 ms without packet loss. Camera streams are encoded with H.264~\cite{h264} and transmitted to the server over WebRTC~\cite{webrtc}. For server-to-client delivery, the server and client are connected over 1 Gbps Ethernet, on which we emulate an RTT of 40 ms without packet loss. We implement a UDP-based transport with a sender-side pacing rate of 250 Mbps and NACK-based retransmission. The client does not employ a jitter buffer.

\noindent\textbf{Dataset.} We evaluate~\sysname~on the CMU PanopticStudio dataset~\cite{panoptic}, a publicly available multi-view capture dataset that has been widely used in volumetric streaming. Among its sequences, we employ \textit{Mafia} which consists of around 18,000 frame sets. The sequence \textit{Mafia} includes multiple people and exhibits view-dependent occlusions, forming dynamic scenes whose geometry changes substantially.

\noindent\textbf{Metrics.} We measure end-to-end latency to assess whether each system supports live interactive streaming. We further evaluate quality at two levels. At the image level, we report SSIM~\cite{ssim} on the frames rendered at the client. At the geometry level, we report the intersection-over-union (IoU) of silhouette between the rendered reconstruction and the reference view, which captures how much of the surface is missing or spuriously reconstructed when a frame set is committed before all views arrive. We use SAM3~\cite{sam3} to extract the person silhouette from each reference image.

\noindent\textbf{Baseline.} To evaluate the end-to-end performance of~\sysname, we compare it against static and adaptive systems.
\begin{itemize}
    \item \textit{Fixed-$\delta$} waits for a fixed interval before committing reconstruction. We evaluate with intervals of $\delta \in \{40, 65, 100\}$ ms. The 40-ms setting represents an aggressive policy that keeps the waiting time close to a single frame interval at 30 fps, whereas the 100-ms setting waits long enough to include nearly all views and thereby minimizes quality loss. To achieve an optimal latency-quality trade-off, we set the commitment threshold to 65 ms, which precisely corresponds to the p99 of the fusion-ready spread observed in our system.
    \item \textit{Event-$\delta$} starts with a waiting interval and resets the timer whenever an additional view becomes available. Specifically, after each newly available view, it waits another $\delta \in \{15, 25, 40\}$ ms from that fusion-ready time---the extension is not accumulated across frames. This policy captures a simple adaptive strategy that continues waiting while views are ready in close succession.
    \item \textit{MetaStream}~\cite{metastream} adaptively adjusts the wait time $U$ every $g$ frame sets by the median difference between the observed view arrival spread and the current $U$, exploiting the temporal correlation of network conditions. It thus increases $U$ when the observed spread has typically exceeded the current $U$, and decreases it otherwise. Following the original implementation, we set $g=30$.
\end{itemize}
To the best of our knowledge, no existing system transmits late views after a frame set has been committed. MetaStream discards views arriving after $U$, and we implement the remaining baselines to behave identically.

\vspace{-0.1in}
\subsection{Overall Performance of~\sysname}

In this section, we evaluate the end-to-end performance of~\sysname, focusing on whether it satisfies the stringent latency requirement of live volumetric streaming while effectively exploiting its pipelined design. Specifically, we first examine the end-to-end latency--quality trade-off under both average and tail view-availability spreads, comparing it against baselines in~\cref{subsec:eval_setup}. We then break down the end-to-end latency to show how early base commitment and late view incorporation overlap with transmission, client-side processing, and rendering, quantifying how this pipelining keeps the displayed result within the target latency budget.

\begin{figure}
    \centering
    \includegraphics[width=1\linewidth]{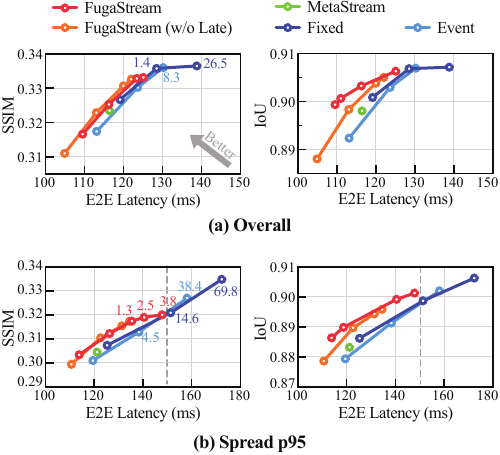}
    \vspace{-0.05in}
    \caption{Latency--quality trade-off of~\sysname~and the baselines. (a) Average performance across all frame sets, (b) performance under the p95 fusion ready spread, where late-arriving views impose larger latency penalties.}
    \label{fig:e2eperf}
\end{figure}

\noindent\textbf{Latency--quality trade-off.} First, we evaluate the end-to-end latency--quality trade-off of~\sysname~and the baselines. Figure~\ref{fig:e2eperf}(a) reports the average E2E latency and rendering quality across frame sets, while Figure~\ref{fig:e2eperf}(b) focuses on the p95 spread of view availability condition, where late views impose a significantly greater latency cost. We report both SSIM and IoU to capture rendered image quality. The number next to each point denotes the percentage of frame groups that fall within a run of three or more consecutive groups violating the latency budget; we annotate only values exceeding 1\%. \sysname~w/o late represents the base representation before any late view is incorporated.

To evaluate all approaches across a broad spectrum of latency regimes, we adjust their respective latency-control parameters. For the Fixed and Event baselines, we sweep their server-side waiting intervals to span different latency budgets. Meanwhile, MetaStream is evaluated at only a single operating point because its reliance on the median arrival time makes its latency not tunable. In contrast, for~\sysname, varying the latency penalty weights---\{0.1, 0.2\} for $\lambda$ and \{50, 100\} for $\tau$---continuously navigates the latency--quality trade-off space, enabling a fair comparison across equivalent latency bounds.

As shown in Figure~\ref{fig:e2eperf}(a),~\sysname~achieves a much better latency--quality trade-off compared to the baselines overall. By committing the base early and incorporating late view updates afterward,~\sysname~retains the latency benefit of early commitment while recovering quality from views that become available later. Specifically,~\sysname~without late view refinement reveals that ABCC triggers significantly earlier than all baselines at comparable quality levels. This confirms that ABCC effectively avoids severe wait time penalties by adaptively terminating the wait for delayed frames unlikely to arrive promptly. In contrast, Fixed improves quality only by waiting longer for additional views, directly increasing latency without notable quality gain at Fixed-100 with 26.5\% exceeding the latency budget. Meanwhile, although MetaStream and Event adjust their commitment timing, without considering the spread in view availability, they fail to reflect the online dynamics, incurring extra latency penalties. Finally, as~\sysname~incorporates late views during client waiting, the reconstruction quality is improved with slightly increased latency---still meeting the 150 ms requirement. Crucially, because late updates prioritize filling novel geometry over refining existing base colors, overall SSIM yields marginal gains while IoU increases substantially. This directly confirms that our novelty extraction mechanism successfully selects and transmits only meaningful data missing from the base commitment.

This advantage of~\sysname~becomes more pronounced under the p95 view availability spread in Figure~\ref{fig:e2eperf}(b). While~\sysname~without late views remains in the low-latency regime by suppressing excessive wait times, enabling client-side render timing control allows late views to be continuously integrated until the application target delay is met. Consequently,~\sysname~delivers the highest reconstruction quality among all schemes while strictly satisfying the 150 ms tail latency budget. In contrast, Fixed and Event fail to satisfy the latency constraint, whereas MetaStream meets it by discarding late views, directly compromising final reconstruction quality.

\begin{figure}
    \centering
    \includegraphics[width=1\linewidth]{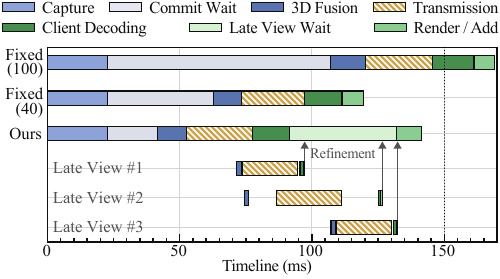}
    \caption{End-to-end timeline breakdown. Late Views \#1--\#3 become available after base commitment and are processed in a pipelined manner.}
    \label{fig:e2ebw}
    \vspace{-0.1in}
\end{figure}

\noindent\textbf{Pipelined execution.} We next examine how~\sysname~reduces end-to-end latency across the pipeline. Figure~\ref{fig:e2ebw} breaks down the execution timeline, comparing Fixed-40 and Fixed-100 against~\sysname. For~\sysname, Late Views \#1--\#3 denote late views that become available after base commitment and are processed as separate updates. The Fixed-100 baseline includes more views but pushes the overall pipeline beyond the 150 ms latency budget, whereas the short wait Fixed-40 commits earlier by excluding views that are not yet ready.

\sysname~commits the base early and immediately advances it through server 3D fusion, transmission, and client decoding. During the remaining latency budget, late views are independently transmitted and processed as shown in Figure~\ref{fig:e2ebw} (Late Views~\#1--\#3). Their processing overlaps with the downstream progression of the base instead of waiting for the base pipeline to complete. This parallel processing allows the fusion-ready delay of late views to be pipelined with the computationally intensive base 3D fusion. For example, by the time the base is fully decoded, Late View \#1 is already queued for immediate integration, making its initial delay completely transparent to the client. Because incorporating these queued updates requires only a lightweight point-wise operation, multiple late views can contribute seamlessly before the final render. This pipelined execution uses the remaining latency budget to recover information from late views while keeping the final display within the 150 ms target. As a result,~\sysname~preserves the latency benefit of early commitment while continuing to improve the representation before rendering.

\begin{figure*}[t]
    \centering
    \includegraphics[width=\linewidth]{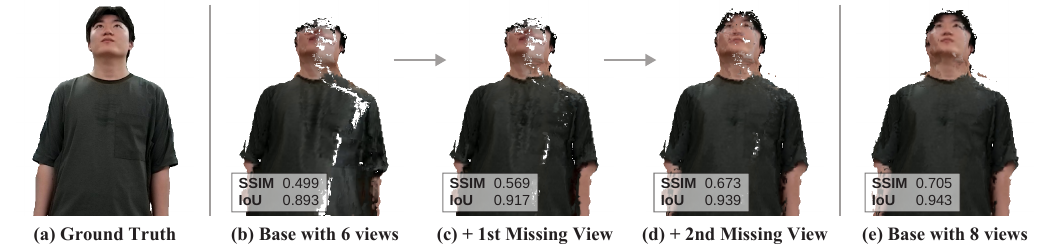}
    \vspace{-0.2in}
    \caption{Qualitative example of base-referenced late view completion: two late views progressively refine a six-view base toward the eight-view reconstruction.}
    \label{fig:real_qual}
\end{figure*}

\begin{figure}
    \centering
    \includegraphics[width=1\linewidth]{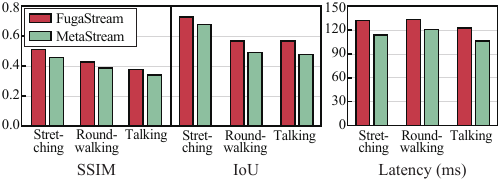}
    \vspace{-0.1in}
    \caption{Real-world testbed evaluation across three captured sequences---\textit{Stretching, Round-walking} and \textit{Talking}.}
    \label{fig:real}
    \vspace{-0.31in} %
\end{figure}

\noindent\textbf{Real-world evaluation.} We further evaluate~\sysname~on our real-world testbed using the hardware described in~\cref{sec:motivation} and Table~\ref{tab:real-world_testbed}. Note that the Orbbec Femto Mega uses the same depth camera module as the Azure Kinect DK~\cite{azure_kinect_dk}, which has been widely adopted in volumetric systems~\cite{livo,livevv,blendr,characterizing}. Each RGB-D camera transmits H.264-encoded color and 16-bit Y16 depth streams using the Orbbec SDK. The server-to-client network follows the same configuration as in~\cref{subsec:eval_setup}. We evaluate three captured sequences---\textit{Stretching}, \textit{Roundwalking}, and \textit{Talking}---with 1,800 frame sets per sequence. The sequences respectively represent a single-person dynamic scene, a four-person dynamic scene, and a four-person relatively static scene.

We first present a qualitative example from one of the real-world traces (Stretching) in Figure~\ref{fig:real_qual}, visualizing how~\sysname~progressively evolves an early committed base into a refined reconstruction as late views become available. (a) shows the ground truth, while (b) shows the base reconstruction formed from the first six available views. The limited coverage of the six-view base leaves several surface regions missing from the initial reconstruction, and this incompleteness is also reflected in the low SSIM and IoU in (b). (c) and (d) show our base-referenced late view completion after each of the two remaining late views is received and incrementally merged. Each refinement fills in regions absent from the initial base, as reflected by the progressive improvements in both SSIM and IoU. After incorporating both late views, the reconstruction in (d) approaches the model built jointly from all eight views in (e).

\begin{figure}
    \centering
    \includegraphics[width=1\linewidth]{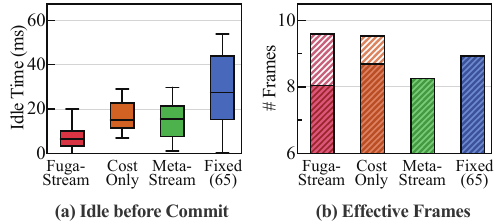}
    \caption{Effectiveness of ABCC under the p95 view-readiness spread. Effective frames are the frames incorporated in the rendered image.}
    \label{fig:bcc_idle}
     \vspace{-0.2in}
\end{figure}

For a quantitative comparison, Figure~\ref{fig:real} compares~\sysname~with MetaStream in terms of SSIM, IoU, and E2E latency. Across all three sequences,~\sysname~consistently achieves higher rendering quality in both SSIM and IoU. MetaStream attains lower latency by committing earlier, but late views that miss its commitment point can no longer contribute to the reconstruction, reducing both perceptual quality and geometric accuracy. In contrast,~\sysname~retains the contribution of these views after early base commitment, yielding higher quality with a modest latency increase. Importantly, its E2E latency remains within the 150 ms target across all sequences. These results show that the latency--quality benefits of~\sysname~persist under real capture, encoding, and transmission dynamics, rather than being limited to the controlled environments used in our earlier experiments.

\subsection{Component-wise Evaluation}
\label{subsec:component-wise}

\noindent\textbf{Idle time and frame utilization.} We examine whether ABCC reduces unnecessary waiting, idle time, without sacrificing frame utilization under severe view-readiness spread. Figure~\ref{fig:bcc_idle}(a) reports the idle time before base commitment, while Figure~\ref{fig:bcc_idle}(b) shows the number of frames the client has received, together with the resulting IoU. We compare our ABCC against cost-only, MetaStream, and \textit{Fixed-65}. Here, cost-only replaces $G_t(\Delta)$ by its upper bound $|\mathcal{M}_g|/N$, \textit{i.e.}, it assumes that every missing view becomes ready within $\Delta$.

ABCC substantially reduces both the magnitude and variability of idle time under the p95 spread. Specifically, its median idle time is roughly 8.5 ms lower than that of the cost-only variant. This difference shows that elapsed waiting time alone is insufficient for an effective commitment decision. The expected coverage gain of the remaining views must also be considered to decide whether to wait longer. This shorter idle time does not reduce frame utilization, as ABCC and the cost-only variant deliver nearly the same total number of frames to the client. Note that the cost-only variant and MetaStream exhibit similar idle times, while MetaStream discards views that become ready after commitment, resulting in lower frame utilization. In contrast,~\sysname~recovers these views through base-referenced late view completion, reducing server-side waiting without permanently sacrificing their contribution.

\begin{figure}
    \centering
    \includegraphics[width=1\linewidth]{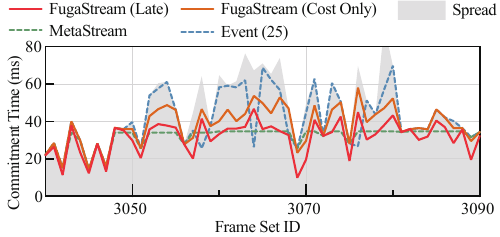}
    \caption{Frame-set-level commitment under varying spread in view availability.}
    \label{fig:bcc_gid}
    \vspace{-0.2in}
\end{figure}

\noindent\textbf{Runtime commitment behavior.} We further examine how ABCC adjusts its commitment timing across consecutive frame sets. Figure~\ref{fig:bcc_gid} plots the commitment time of each scheme over a \textit{Mafia} sequence, together with the observed spread in view availability. We compare~\sysname~against the \textit{cost-only} variant, MetaStream, and Event-25.

\sysname~adapts its commitment time on a frame-set basis without simply following fluctuations in the readiness spread. As waiting progresses, ABCC continuously re-evaluates both the expected contribution of the remaining views and their conditional probability of becoming ready within additional waiting. It therefore stops waiting when the remaining views are unlikely to become ready soon or their expected coverage gain no longer justifies the waiting cost, even under a large spread in view availability. In contrast, cost-only generally waits longer because it cannot distinguish whether the remaining views justify additional delay, while Event reacts strongly whenever a new view becomes ready and exhibits larger fluctuations in commitment time. MetaStream shows the opposite behavior: its median-based adaptation suppresses tail behavior, causing the last-ready views that form the large spread to have little impact on its waiting time and thus to be often discarded. By making each commitment decision from the current state,~\sysname~responds to transient changes without either aggressively tracking individual late views or ignoring tail readiness behavior.

\begin{figure}
    \centering
    \includegraphics[width=\linewidth]{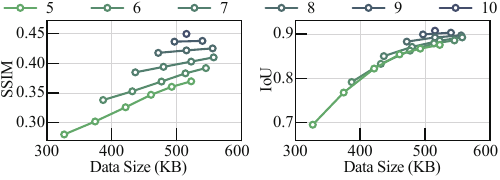}
    \caption{Effect of late view refinement for different base sizes.}
    \label{fig:NF_1}
\end{figure}

\begin{figure}
    \centering
    \includegraphics[width=\linewidth]{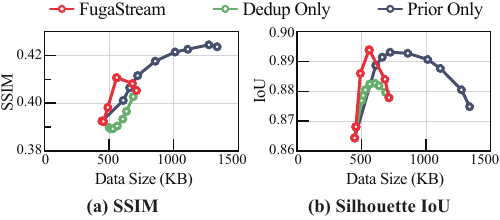}
    \caption{Ablation study of base-referenced late view completion. \textit{Prior Only} considers only novelty score (Eq.~\ref{eq:novelty_score}), while \textit{Dedup Only} maintains only a transient overlap map per frame set.}
    \label{fig:NF_ablation}
\end{figure}

\noindent\textbf{Reconstruction gain from late views.} We examine whether late views provide meaningful reconstruction gains when incorporated through~\sysname's late view refinement path. Out of ten total views, we vary the number of views $B$ included in the base representation from five to nine, and incrementally apply the remaining $M=10-B$ views as late view updates. Each curve fixes a base of $B$ views and extends it one late view at a time, so the leftmost point is the base-only quality and the rise along the curve is the gain attributable to late views alone in Figure~\ref{fig:NF_1}.

The results demonstrate two points. First, late view refinement is beneficial across all base sizes, consistently improving reconstruction quality. Late views supplement surface information not sufficiently captured by base representation, improving SSIM by up to 0.09 and IoU by up to 0.18. Second, late views cannot fully compensate for insufficient base coverage. Even with ten total views, reconstructions initialized from a smaller base perform worse than those initialized from a larger base. This indicates that committing too early imposes a lower ceiling on the final reconstruction quality, motivating the server to wait until a sufficiently large base has been formed.

\noindent\textbf{Effect of novelty extraction.} The core objective of~\sysname's novelty extraction is to drastically reduce transmission overhead while maximally preserving 3D reconstruction quality. To dissect how each mechanism contributes, we compare the integrated novelty extraction against two ablated variants: (i) \textit{Prior Only}, which filters points using only the novelty score against the base representation (Eq.~\ref{eq:novelty_score}), and (ii) \textit{Dedup Only}, which selects points using only the overlap map among the late views (Figure~\ref{fig:NF_ablation}). Each marker in the figure corresponds to a value of the novelty threshold $\theta_p$; Dedup Only, which does not use the score, is instead rate-matched to~\sysname~by drawing the same number of points at random at each threshold.

To isolate the impact of inter-view deduplication among late views, we first compare~\sysname~with the Prior Only baseline. Deduplication is the primary driver of our bandwidth savings. It aggressively restricts updates to a single point per voxel for each frame set, filtering out redundant points and reducing the overall 3D frame size. Specifically, without deduplication, the Prior Only approach consumes 723 KB per 3D frame at its best operating point, which is 29\% larger than 560 KB. As shown in Figure~\ref{fig:NF_1}, this strict spatial deduplication ensures that total bandwidth consumption is determined primarily by the number of views incorporated and only weakly by how they are split between the base and late updates.

However, relying solely on deduplication (Dedup Only) curtails bandwidth at the direct expense of reconstruction quality. Because deduplication permits only one update per voxel, the retained points must each yield meaningful improvements in geometric (IoU) and visual (SSIM) fidelity. Without utility-based prioritization, Dedup Only fails to filter out noise; while it initially constructs the core subject, it subsequently allocates precious bandwidth to noisy exterior regions, causing IoU to degrade. In contrast,~\sysname's priority filtering specifically isolates base-relative novelty. By forwarding only essential updates and preventing noise from preempting critical data,~\sysname~drives a sharp increase in both metrics, avoiding the baseline's degradation.

\begin{figure}
    \centering
    \includegraphics[width=1\linewidth]{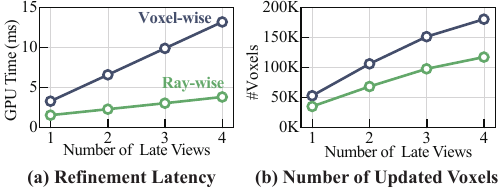}
    \caption{Client-side refinement overhead of ray-wise and voxel-wise TSDF fusion measured at the NVIDIA Jetson Orin NX.}
    \label{fig:NF_raycast}
\end{figure}

Overall,~\sysname's novelty extraction effectively balances geometric recovery with transmission efficiency. By accurately selecting geometric novelty, the system achieves an IoU comparable to the unfiltered baseline, successfully reconstructing the 3D shape. The inherent trade-off lies in color refinement: because deduplication explicitly prevents redundant updates to the already committed base, color enhancements are restricted to the newly incorporated novel regions, slightly capping the absolute peak SSIM. Nevertheless, this precise compromise is what empowers~\sysname~to integrate late-arriving views and preserve geometric fidelity with similar bandwidth overhead.

\noindent\textbf{Base refinement overhead.} Our system requires base refinement to remain lightweight, so that incorporating late views does not compromise the latency benefit of early commitment. We evaluate the computational overhead of the base refinement at the client. Figure~\ref{fig:NF_raycast}(a) compares the GPU computation time of our ray-wise refinement with a voxel-wise alternative as the number of views increases, while Figure~\ref{fig:NF_raycast}(b) reports the number of voxels updated by each method.

As more late views arrive, the latency gap between the two approaches widens: ray-wise refinement remains lightweight, whereas voxel-wise refinement incurs 2.14--3.47$\times$ higher latency. Ray-wise refinement follows only the rays associated with each late view and directly updates the voxels required by the newly observed regions. In contrast, voxel-wise refinement must not only apply the required updates but also access every voxel in the committed base to identify them including voxels irrelevant to the late view. This unnecessary traversal becomes increasingly costly as late views accumulate. By restricting computation to the voxels that actually require refinement, ray-wise refinement makes client-side base updates practical with minimal latency overhead.

\section{Discussion}
\label{sec:Discussion}

\noindent\textbf{Application-aware runtime optimization.} Our current design demonstrates the effectiveness of pipelined volumetric streaming using lightweight runtime information, the history of fusion-ready times and the cost of additional waiting. Nevertheless, further runtime optimization could improve the latency--quality trade-off. For example, applications that can accommodate richer per-frame inference could employ DNN-based predictors to extract complex temporal and scene-dependent patterns for more sophisticated commitment decisions. Other runtime signals, such as network bandwidth, server load, and rendering feedback, could also inform these decisions. Such information and mechanisms can be incorporated in an application-aware manner, with the decision complexity bounded according to latency requirements and the available inference and control budget.

\noindent\textbf{Heterogeneous downstream conditions.} Our current evaluation primarily validates~\sysname~under controlled downstream conditions. In practice, a deployment may serve multiple users simultaneously, while downlink bandwidth can vary over time and across users. Clients may also differ substantially in their available computation resources and application-level latency requirements. These heterogeneous downstream conditions can lead to different preferred operating points between latency, bandwidth consumption, and reconstruction quality. Although our evaluation does not explicitly characterize these scenarios, we expect that the proposed pipeline naturally provides adaptation points for such heterogeneous operating conditions. Specifically,~\sysname's runtime prioritization can be adjusted to control when the base is committed and how much late view information is subsequently delivered, enabling different operating points without fundamentally changing the underlying pipeline.

\noindent\textbf{Bandwidth-aware delivery.} \sysname~currently uses Draco to compress reconstructed geometry, but Draco provides only limited support for adapting the transmitted representation to changing bandwidth conditions. With more advanced volumetric codecs including learned or scalable codecs, we believe that our system can further reduce the size of both the base representation and subsequent late view updates, and adapt to server-to-client network bandwidth. Specifically, the separation between the early delivered base and subsequent late view updates provides a natural opportunity for bandwidth adaptation: under constrained bandwidth, the system can prioritize the base and selectively transmit only essential updates, while additional updates can be delivered as more bandwidth becomes available. We leave such bandwidth-aware adaptation as a valuable direction for future work.

\section{Related Work}
\label{sec:related_work}

\noindent\textbf{Live volumetric streaming.} End-to-end systems for live volumetric streaming divide naturally by the dimensionality of what they put on the network: reconstructed 3D content, or the 2D sensor data from which 3D representation is later reconstructed, at the client. On the 3D side, MetaStream~\cite{metastream} assumes cameras equipped with computation resources, and its server adaptively sets a wait time for each frame set based on the median arrival time of the 2D frames in the preceding set. PointPresence~\cite{pointpresence} has the server run object detection on incoming RGB-D frames and forward only the point cloud inside the bounding box to the edge that decodes them. The 2D side is motivated by a stark asymmetry in codec maturity. FarfetchFusion~\cite{farfetchfusion}, specializing in mobile live telepresence, has the sender extract 2D and 3D landmarks and transmit transformed RGB-D, leaving the receiver to perform fusion and recover a 3D mesh. In LiVo~\cite{livo} for immersive volumetric conferencing, the sender takes the viewer’s frustum from the receiver and transmits only the corresponding RGB and depth streams, packing depth into the Y channel and UV channels with a fixed value so that a video codec carries it. Both branches leave one assumption untouched: the set of views contributing to a frame is fixed at deployment time, and a view that misses its deadline contributes nothing rather than being recovered later.

\noindent\textbf{Multi-view compression.} Orthogonal to streaming system architecture, another body of work addresses the limitation that bandwidth consumption in volumetric streaming scales with the number of cameras. DeltaStream~\cite{deltastream} converts inter-frame change in 2D RGB and depth into 3D motion vectors, then compresses and transmits the delta point cloud against the previous frame together with those vectors. BlenDR~\cite{blendr} packs depth into the Y channel and color into the UV channels, multiplexing a low-resolution UV stream with a Y depth stream for transmission. In ImmerScope~\cite{immerscope}, focusing on inter-view dependency, each camera transmits its frames through a DNN-based encoder, and the server aggregates the multiple views into a single frame that a DNN-based decoder reconstructs. Standard point cloud and voxel codecs such as Draco~\cite{draco} and the MPEG V-PCC/G-PCC families~\cite{vpcc_gpcc}, meanwhile, produce bitstreams that remain decodable for general-purpose 3D compression. Our system’s position with respect to this line of work is that we exploit the same redundancy not to shrink each view before it is sent, but to recover, from an early-committed base within the same frame set, the views that miss their deadline.

\section{Conclusion}
\label{sec:conclusion}

In this paper, we present \sysname, a pipelined volumetric streaming system that mitigates latency caused by spread in view availability without permanently sacrificing information from late views. Instead of stalling until all views are ready, \sysname~adaptively commits a 3D base representation and incrementally refines it as late views become available. Its adaptive base commitment controller balances the expected geometric gains from additional views against the latency cost of waiting, while base-referenced late view completion identifies and delivers only novel geometry not already represented in the base. We evaluate \sysname~using both a standard dataset and a real-world testbed to demonstrate its effectiveness under diverse view-availability patterns. The results show that \sysname~substantially reduces end-to-end latency while preserving high rendering quality, with lightweight late view updates that retain the benefits of early commitment.~\sysname~demonstrates that committing and delivering a useful base before all views are ready, while incorporating late views afterward, provides a practical path toward low-latency volumetric streaming under uncertain view availability.

\bibliographystyle{ACM-Reference-Format}
\bibliography{reference}

\end{document}